\documentclass[12pt]{article}

\usepackage{graphicx}
\begin{document}
\begin{center}
{\Large Relativistic, Causal Description of Quantum Entanglement
and Gravity}
\vskip 0.3 true in {\large J. W. Moffat}
\date{} \vskip 0.3 true in {\it Department of Physics,
University of Toronto, Toronto, Ontario M5S1A7, Canada, \vskip
0.1 true in and \vskip 0.1 true in Perimeter Institute for
Theoretical Physics, Waterloo, Ontario N2J2W9, Canada.}

\end{center}

\begin{abstract}%
A possible solution to the problem of providing a spacetime
description of the transmission of signals for quantum entangled
states is obtained by using a bimetric spacetime structure, in which
quantum entanglement measurements alter the structure of
the classical relativity spacetime. A bimetric gravity
theory locally has two lightcones, one which describes
classical special relativity and a larger
lightcone which allows light signals to communicate quantum
information between entangled states, after a measurement device
detects one of the entangled quantum states. This theory would
remove the tension that exists between macroscopic classical, local gravity
and macroscopic nonlocal quantum mechanics.

\end{abstract}
\vskip 0.2 true in e-mail: john.moffat@utoronto.ca

\vskip 0.3 true in

\section{\bf Introduction}

One of the most important features of quantum mechanics is the
Einstein-Podolsky-Rosen~\cite{Einstein,Schrodinger} effect, in
which strong correlations are observed between presently
noninteracting particles that have interacted in the past. The
problem of understanding the consequences of the EPR effect is
still controversial~\cite{Laloe}. Experiments on {\it entangled}
particle states have verified the `nonlocal' nature of quantum
mechanics~\cite{Gisin}. One disturbing feature of the standard
interpretation of quantum mechanics is that the nonlocal nature
of the entanglement process has been divorced from our common
intuitive ideas about spacetime events and causality. The
standard interpretation asserts that for photons (or electrons)
positioned at A and B, separated by a spacelike distance, there
is no exchange of classical information and superluminal signals
between A and B are impossible according to special relativity.
With the advent of the possibility of constructing quantum
computers and performing `teleportation' experiments, the whole
issue of the spacetime reality of the EPR process becomes more
problematic.

There exists also the fundamental puzzle that contemporary quantum mechanics is
nonlocal on a macroscopic level, whereas gravitation described by Einstein's general
relativity (GR) is a strictly local macroscopic theory. This causes a tension
to exist between the two fundamental pillars of modern physics~\cite{Chiao}.

We can adopt three possible positions:
\begin{enumerate}
\item There is no problem. Quantum mechanics is nonlocal and we should accept that
there is no possible causal phenomenon associated with a space and time
interpretation of entanglement as dictated by classical special relativity and
Bell's inequality~\cite{Bell}.

\item Quantum mechanics should be altered in some way to bring about a causal, space
and time description of quantum mechanics.

\item Classical spacetime locally
described by a flat, Minkowskian metric with one light cone, is
not adequate to explain the physics of quantum entanglement.
The standard, classical description of spacetime must be
extended when quantum mechanical systems are measured.
\end{enumerate}

According to (1) quantum entanglement is a {\it purely quantum phenomenon} and
classical concepts associated with causally connected events in space are absent.
This is the point of view advocated by practitioners of standard quantum mechanics.
One should abandon any notion that physical space plays a significant role for
distant correlations of entangled quantum states.  For those who remain troubled by
this abandonment of a causal spatial connection between entangled
states, it is not clear how attempting to change quantum mechanics would help
matters. This leaves the possibility (3) that classical special relativity is too
restrictive to allow for a complete spacetime, causal description of quantum
entanglement.

In the following, we adopt position (3) and propose a
scenario based on a `bimetric' description of spacetime. This kind of construction
has been successful in cosmology, in which it provides an alternative to the
standard inflationary
cosmologies~\cite{ClaytonMoffat,ClaytonMoffat2,ClaytonMoffat3}. In the
present application of the bimetric description of spacetime to the
quantum entanglement problem, we picture that a quantum mechanical metric
frame is related to a gravitational metric frame by the gradients of a
scalar field $\phi$.  The gravitational metric describes {\it locally} a
Minkowski light cone with constant speed of gravitational waves
(gravitons) $c_g$, while the quantum mechanical metric describes locally a
different light cone with an increased speed of light $c>c_0$, where $c_0$
is the currently observed speed of light. When the dimensionless parameter
$\gamma=c/c_g=1$, spacetime is described locally by a flat Minkowski metric
with one fixed lightcone and we can choose units such that $c=c_g=c_0=1$.

The amount of entanglement of a quantum mechanical bipartite
system is given by the density matrix of its von Neumann
entropy. For a pure non-entangled state, the speed of transmitted
signals travels with the standard classical, special relativity
value $c_0$, but for entangled states, quantum mechanical
superluminal signals can travel in the quantum mechanical metric
frame, thereby providing a Lorentz invariant {\it
spacetime} description of quantum entanglement phenomena. When an
entangled quantum system suffers decoherence due to
environmental effects, the system rapidly becomes a
classical one with the spacetime structure determined by the
gravitational metric $g_{\mu\nu}$, corresponding locally to Minkowski spacetime with
a single lightcone.

\section{\bf Bimetric Gravity Theory}

We postulate that four-dimensional spacetime is described by the
bimetric structure~\cite{ClaytonMoffat}:
\begin{equation}
\label{bimetricscalar}
{\hat g}_{\mu\nu}=g_{\mu\nu}+\alpha\partial_\mu\phi\partial_\nu\phi,
\end{equation}
where $\alpha>0$ is a constant with dimensions of $[{\rm length}]^2$ and we
choose the scalar field $\phi$ to be dimensionless. The metric
${\hat g}_{\mu\nu}$ is called the matter quantum mechanical metric, while
$g_{\mu\nu}$ is the gravitational metric. The scalar field $\phi$ belongs to the
gravitational sector. We choose the signature of flat spacetime to be described by
the Minkowski metric $\eta_{\mu\nu}={\rm diag}(1,-1,-1,-1)$. The inverse metrics
${\hat g}^{\mu\nu}$ and $g^{\mu\nu}$ satisfy
\begin{equation}
{\hat g}^{\mu\alpha}{\hat g}_{\nu\alpha}={\delta^\mu}_\nu,\quad
g^{\mu\alpha}g_{\nu\alpha} ={\delta^\mu}_\nu.
\end{equation}
We assume that only non-degenerate values of ${\hat g}_{\mu\nu}$ with
${\rm Det}({\hat g}_{\mu\nu})\not= 0$ correspond to physical spacetime.

The action is given by
\begin{equation}
\label{scalaraction}
S=S_G[g]+S_\phi[g,\phi]+{\hat S_M}[{\hat g},\psi^I],
\end{equation}
where
\begin{equation}
S_G[g]=-\frac{1}{\kappa}\int dt d^3x\sqrt{-g}(R[g]+2\Lambda),
\end{equation}
and
\begin{equation}
S_\phi[g,\phi]=\frac{1}{\kappa}\int
dtd^3x\sqrt{-g}\biggl(\frac{1}{2}g^{\mu\nu}\partial_\mu\phi\partial_\nu\phi-V(\phi)\biggr).
\end{equation}
Moreover, the matter stress-energy tensor is
\begin{equation}
{\hat T}^{\mu\nu}=-\frac{2}{\sqrt{-{\hat g}}}\frac{\delta {\hat S}_M}{\delta {\hat
g}_{\mu\nu}}, \end{equation}
$\kappa=16\pi G/c_0^4$, $\Lambda$ is the cosmological
constant and $\psi^I$ denotes matter fields. We have
constructed the matter action ${\hat S}_M$ using the matter quantum mechanical
metric ${\hat g}_{\mu\nu}$. The stress-energy tensor for
the scalar field $\phi$ is given by
\begin{equation}
T_{\phi}^{\mu\nu}=\frac{1}{\kappa}\biggl[g^{\mu\alpha}g^{\nu\beta}\partial_\alpha\phi\partial_\beta\phi
-\frac{1}{2}g^{\mu\nu}g^{\alpha\beta}\partial_\alpha\phi\partial_\beta\phi+g^{\mu\nu}V(\phi)\biggr].
\end{equation}

An alternative choice of bimetric structure
is~\cite{ClaytonMoffat2,ClaytonMoffat3}:
\begin{equation}
\label{bimetricphoton}
{\hat g}_{\mu\nu}=g_{\mu\nu}+\alpha\psi_\mu\psi_\nu,
\end{equation}
where $\psi_\mu$ is a vector field and
\begin{equation}
F_{\mu\nu}=\partial_\mu\psi_\nu-\partial_\nu\psi_\mu
\end{equation}
is the field strength tensor. In the following, we shall consider only the
bimetric tensor-scalar spacetime structure defined by (\ref{bimetricscalar}).

Variation of the action $S$ in (\ref{scalaraction}) gives the
field equations
\begin{equation}
G^{\mu\nu}=\frac{\kappa}{2}(s{\hat
T}^{\mu\nu}+T_{\phi}^{\mu\nu})+\Lambda g^{\mu\nu},
\end{equation}
\begin{equation}
\label{scalarmotion}
\nabla_\mu\nabla^\mu\phi+V'(\phi)-\kappa s\alpha{\hat
T}^{\mu\nu}{\hat\nabla}_\mu{\hat\nabla}_\nu\phi=0,
\end{equation}
where $G^{\mu\nu}=R^{\mu\nu}-(1/2)g^{\mu\nu}R$ is the Einstein
tensor and $s=\sqrt{-{\hat g}}/\sqrt{-g}$. Moreover, $\nabla_\mu$
and ${\hat\nabla}_\mu$ denote the covariant differential
operators associated with $g_{\mu\nu}$ and ${\hat g}_{\mu\nu}$,
respectively, $V(\phi)$ is the potential for the field $\phi$
and
$V'(\phi)=\partial V(\phi)/\partial\phi$. For a free scalar field $\phi$ the potential will be
$V(\phi)=\frac{1}{2}m^2\phi^2$, where $m$ is the mass of the particle
associated with the scalar field $\phi$.  The energy-momentum
tensor ${\hat T}^{\mu\nu}$ satisfies the conservation laws
\begin{equation}
{\hat\nabla}_\nu\biggl[\sqrt{-{\hat g}}{\hat
T}^{\mu\nu}\biggr]=0. \end{equation}

\section{\bf Bimetric Special Relativity and Quantum Mechanics}

The local special relativity metric is given by $g_{\mu\nu}=\eta_{\mu\nu}$ with
\begin{equation}
\label{SRM}
ds^2\equiv\eta_{\mu\nu}dx^\mu dx^\nu=c_g^2dt^2-(dx^i)^2,
\end{equation}
where ($i,j=1,2,3$). The quantum mechanical metric for the choice
(\ref{bimetricscalar}) is
\begin{equation}
\label{QMM}
d{\hat s}^2\equiv {\hat g}_{\mu\nu}dx^\mu dx^\nu=(\eta_{\mu\nu}+\alpha\partial_\mu\phi
\partial_\nu\phi)dx^\mu dx^\nu.
\end{equation}
The latter can be written as
\begin{equation}
d{\hat s}^2=c_0^2dt^2\biggl(1+\frac{\alpha}{c_0^2}{\dot\phi}^2\biggr)-(\delta_{ij}
-\alpha\partial_i\phi\partial_j\phi)dx^idx^j,
\end{equation}
where $\dot\phi=d\phi/dt$. We see that the speed of light in the quantum
mechanical metric
is space and time dependent. If we assume that $\partial_i\phi\approx 0$, then we have
\footnote{In GR, with one local lightcone, we can always perform a
diffeomorphism transformation to remove the time dependence of $c$. This correponds
to being able to choose units in which rigid ruler and clock measurements yield
$\Delta c/c=0$.  In the bimetric gravity theory, we cannot simultaneously remove the
time dependence of $c$ and $c_g$ by performing a diffeomorphism
transformation. If we choose $c_g$ to be constant, then the time
dependence of $c$ is non-trivially realized.}
\begin{equation}
c(t)=c_0\biggl(1+\frac{\alpha}{c_0^2}\dot\phi^2\biggr)^{1/2}.
\end{equation}

The null cone equation $ds^2=0$ describes gravitational wave signals
moving with the constant measured speed $c_g$, whereas $d{\hat s}^2=0$ cannot be
satisfied along the same null cone lines, but determines an expanded null
cone with the speed of light $c > c_0$.  The bimetric
null cone structure is described in Fig.1.
\vskip 0.2 in
\begin{center}
\includegraphics[width=3.0in,height=3.0in]{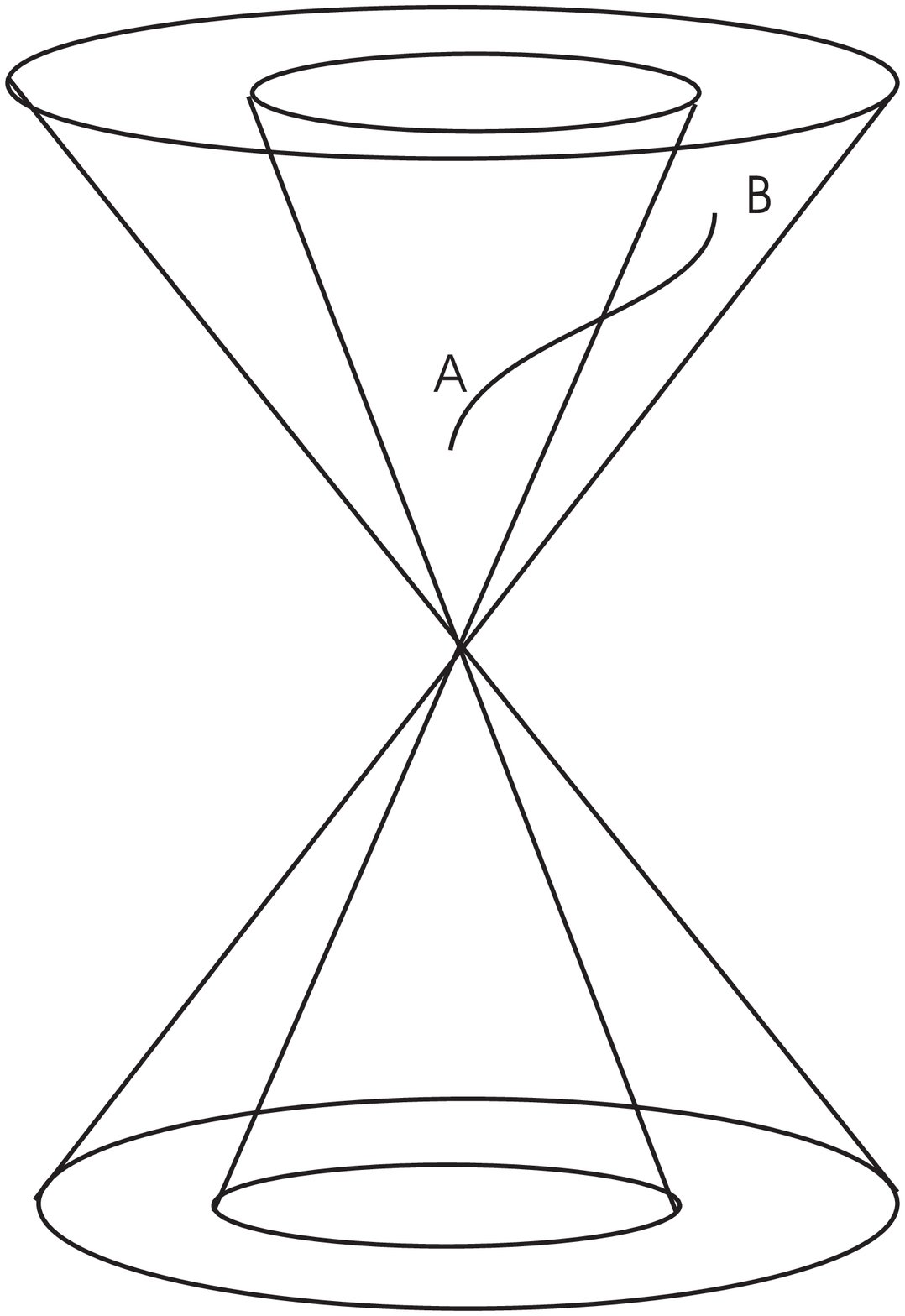}
\end{center}
\vskip
0.1 in
\begin{center}
Fig. 1. Bimetric light cones showing the timelike communication
path between the two entangled states at A and B in the quantum mechanical metric
${\hat g}_{\mu\nu}$.
\end{center}
\vskip 0.1 true in

Both metrics (\ref{SRM}) and (\ref{QMM}) are separately
invariant under {\it local} \footnote{We consider a small patch of
spacetime in which $g_{\mu\nu}\approx \eta_{\mu\nu}$, so that we can
restrict our attention to local Lorentz transformations.} Lorentz
transformations
\begin{equation} x^{'\mu}={\Lambda^\mu}_\nu x^\nu,
\end{equation}
where ${\Lambda^\mu}_\nu$ are constant matrix coefficients which satisfy
the orthogonality condition
\begin{equation}
{\Lambda^\mu}_\nu{\Lambda_\mu}^\sigma={\delta_\nu}^\sigma.
\end{equation}

The `physical' matter metric ${\hat g}_{\mu\nu}$ describes the
geometry in which quantum matter propagates and interacts. Because all
quantum matter fields are coupled to the same metric ${\hat g}_{\mu\nu}$ in
the same way, the weak equivalence principle is satisfied in this metric
frame. However, because ${\hat g}_{\mu\nu}$ does not couple to matter in
the same way as in GR unless $\phi=0$, the strong equivalence principle is
not satisfied.

If we choose a spacelike surface for each individual metric ${\hat
g}_{\mu\nu}$ and $g_{\mu\nu}$, then a Cauchy initial value solution of the
field equations for given initial data on these spacelike surfaces can be
obtained, which preserves the causal evolution of the fields and field
equations~\cite{ClaytonMoffat3}. If we consider a vector field $V_\mu$,
which is timelike (or null) with respect to both metrics ${\hat
g}_{\mu\nu}$ and $g_{\mu\nu}$, then an observer would see a normal, causal
ordering of timelike events. However, we could picture a situation in which
an observer at $A$, in Fig.1, observes that the vector field $V_\mu$ is
timelike with respect to the smaller light cone, but is spacelike with
respect to the second larger lightcone. In this sense, causality in the
bimetric gravity theory is observer dependent. Thus, by performing a
Lorentz transformation to a boosted frame in the second, larger light
cone, events connected by a `superluminal' quantum information signal
would appear to be acausal with respect to the observer at $A$, i.e. the
emission of a light signal would appear to take place later than its
reception. However, In the `physical' matter metric frame described by
${\hat g}_{\mu\nu}$, an observer at $B$ would see a proper timelike and
causal ordering of events, provided the observer is restricted to
observing a timelike vector field. Thus, the emission of quantum matter
information signals would be causal with respect to the observer at $B$ in
the metric frame described by ${\hat g}_{\mu\nu}$.

In classical physics, information is communicated through space with a limiting speed
$c=c_0$. Information can affect events only in the forward light
cone. We are concerned with the transmission of `quantum information' (QI), which is
transported through space at some speed $v_{QI}$. Let us consider a smallest cone
with $\theta_{QI}$ the angle of the cone to the vertical. Then $v_{QI}=0$
corresponds to $\theta_{QI}=0$, $v_{QI}=c_0=c_g$ to $\theta_{QI}=\pi/4$, and $v_{QI}
= \infty$ corresponds to $\theta_{QI} = \pi/2$~\cite{Garisto}.
We can also define an inverse speed of QI:
\begin{equation}
w_{QI}=\cot{\theta_{QI}}=\biggl(\frac{c_0}{v_{QI}}\biggr).
\end{equation}
Then, $v_{QI}=c_g=c_0$ and $\infty$ correspond to $w_{QI}=1$ and $0$, respectively.
The case $w_{QI}=1$ is related to the classical special relativity metric frame with
the constant values $c=c_0=c_g$, while $0\le w_{QI}\le 1$ corresponds to the quantum
mechanical metric lightcone swept out by the time dependence of $c=c(t)$ for
$\alpha\not=0$.

We require a local relativistic description of quantum mechanics. To
this end, we introduce the concept of a general spacelike surface in
Minkowski spacetime, instead of the flat surface $t={\rm constant}$. We
demand that the normal to the surface at any point $x$,
$n_\mu(x)$, be time-like: $n_\mu(x)n^\mu(x) > 0$.  We denote a spacelike surface by $\sigma$.
A local time $t({\vec x})$ is assigned, so that in the limit that the surface becomes
plane, each point has the same time $t={\rm constant}$. We can
now define the Lorentz invariant functional derivative
$\delta/\delta\sigma(x)$ and the Tomonaga-Schwinger
equation ~\cite{Tomonager,Schwinger,Schweber}
\begin{equation}
\label{Tomonaga-Schwinger}
i\hbar c_0\frac{\delta\psi(\sigma)}{\delta\sigma(x)}={\cal H}_{\rm
int}(x)\psi(\sigma),
\end{equation}
where
\begin{equation}
H_{\rm int}(t)=\int d^3x{\cal H}_{\rm int}(x),
\end{equation}
is the Hamiltonian operator in the interaction representation, and ${\cal
H}_{\rm int}(x)$ is the Lorentz invariant Hamiltonian density. Eq.
(\ref{Tomonaga-Schwinger}) is a relativistic extension of the
Schr\"odinger equation
\begin{equation}
i\hbar\partial_t\psi(t)=H(t)\psi.
\end{equation}

It is essential that the domain of variation of $\sigma$ is restricted by
the integrability condition
\begin{equation}
\label{integr}
\frac{\delta^2\psi(\sigma)}{\delta\sigma(x)\delta\sigma(x')}-
\frac{\delta^2\psi(\sigma)}{\delta\sigma(x')\delta\sigma(x)}=0.
\end{equation}
This equation in turn implies that
\begin{equation}
\label{comm}
[{\cal H_{\rm int}}(x),{\cal H_{\rm int}}(x')]=0,
\end{equation}
for $x$ and $x'$ on the spacelike surface $\sigma$. In quantum field
theory, it is usual to work in the interaction picture, so that the
invariant commutation rules for the field operators automatically guarantee
that (\ref{comm}) is satisfied for all interacting fields with local
nonderivative couplings.

The Tomonaga-Schwinger equation evolves unitarily in the
special relativity metric $g_{\mu\nu}=\eta_{\mu\nu}$ with $\alpha=0$, before a
measurement of a quantum mechanical state is performed and before
the collapse of the state wave function. After a measurement is performed on
a quantum state and $\alpha$ is `switched on', then depending upon the spatial
distance of the causal communication between two entangled states $A$ and $B$, and
the size of $\alpha$ required to make $A$ and $B$ timelike separated, the
Tomonaga-Schwinger equation and its non-relativistic counterpart -- the
Schr\"odinger equation -- will have a spacelike region outside the quantum
mechanical metric lightcone to evolve unitarily. For $\alpha\rightarrow\infty$
this spacelike region will shrink to zero and $A$ and $B$ will be timelike separated
by an infinite spatial distance and $v_{QI}=\infty$.

We introduce the concept of a locally Lorentz invariant density matrix:
\begin{equation}
\rho(\sigma(x))=\vert\psi(\sigma)\rangle\langle\psi(\sigma)\vert.
\end{equation}
The density matrix operator $\rho(\sigma)$ satisfies the invariant
Heisenberg equation of motion
\begin{equation}
i\hbar c_0\frac{\delta\rho(\sigma)}{\delta\sigma}
=[{\cal H_{\rm int}}(x),\rho(\sigma)].
\end{equation}

A local relativistic measure of
the entanglement of a bipartite quantum state is given by
\begin{equation}
\label{relaentropy}
S(\rho_m)(\sigma)=-{\rm Tr}_m\rho(\sigma)\log\rho(\sigma),
\end{equation}
where $S(\rho_m)(\sigma)$ is the relativistic entropy of the
subsystems A and B. Moreover, $\rho_A={\rm
Tr}_B\vert\psi\rangle\langle\psi\vert$ is the reduced density matrix
obtained by tracing the whole system's pure state density matrix
$\rho_{AB}=\vert\psi\rangle\langle\psi\vert$ over A's degrees of freedom,
while $\rho_B={\rm Tr}_A\vert\psi\rangle\langle\psi\vert$ is the partial
trace over B's degrees of freedom.  In the non-relativistic limit, (\ref{relaentropy}) reduces to
the pure bipartite entropy measure of entanglement~\cite{Bennett,Wootters}.
\footnote{We consider in the following only the simple physical system of two
photons (or two electrons). The entanglement measure for a mixed state and a
multiparticle state is controversial and no consensus has been reached on how to
define it.}

For a non-entangled state for which the
$\psi(\sigma)$ can be expressed as a tensor product
$\psi_A(\sigma)\otimes\psi_B(\sigma)$, we have $S(\rho)(\sigma)=0$, ${\hat
g}_{\mu\nu}= g_{\mu\nu}$ and there is no signal transmitted between the bipartite
states $A$ and $B$ by light signals associated with the
special relativity metric (\ref{SRM}) light cone with $c_g=c_0$, for they are now
spacelike separated.
 
For an entangled state $\alpha\not= 0$ the
spacetime is described by the local metric (\ref{QMM}). Since the
speed of light can become much larger in the quantum mechanical metric
frame, it is now possible to transmit signals at `superluminal'
speeds without violating the spatial causality notions that prevail in
the familiar classical special relativity frame. In
this way, by introducing a bimetric spacetime, we have
incorporated the notions of spacetime events and causality for
quantum mechanical entangled states.

\section{\bf Conclusions}

An observer who detects
a quantum mechanical system with some measuring device and subsequently
observes an entangled state
with $\alpha\not=0$ in the quantum mechanical metric frame, will also observe an
exchange of quantum information with the spatially distant other component of the
entangled state.  In the quantum mechanical metric frame, the speed of the light
signal $c$ emitted, say, at a counter at A and received at a distant counter B will
be {\it finite} but large compared to the measured value of the speed of light $c_0$
in the classical, local gravitational metric frame, and A and B are no longer
`spacelike' separated.

A possible experiment to verify the bimetric spacetime scenario would be to attempt
to measure a decrease in the communication signals
associated with the solutions of the fields
$\phi$ with $\alpha\not=0$. Since electromagnetic (photon) communication can take
place over large cosmological distances in the universe, then given enough
transmission energy the two entangled states $A$ and $B$ could communicate at large
spatially separated distances without an observer detecting a significant decrease in
the amount of entanglement and correlation of A and B. The current correlation
experiments have a separation distance $\sim$ 11 kilometres and it is estimated in a
preferred frame that $c\sim 10^4c_0$~\cite{Gisin}. It is possible that when
experiments are performed at larger spatial separations, a dilution of the
correlation between entangled states will be observed. Until then, we cannot exclude
our alternative proposal, based on an extended spacetime structure using a bimetric
geometry.

In our bimetric scenario, the Bell inequalities would not exclude
our extension of special relativity theory and gravitation
theory. The observed experimental violations of Bell's inequality tell us that the
assumption of classical locality, based on one metric and one
Minkowski light cone, is not compatible with quantum mechanics.
But the observed violation of Bell's inequality does not exclude the
extended concept of locality possessed by the bimetric
two-lightcone structure and quantum mechanics.

In our generalized bimetric gravity theory, we no longer have the
peculiar tension that exists between GR and
quantum mechanics, caused by classical gravity theory being a
strictly local theory at macroscopic distances (equivalence
principle) and the nonlocal behavior of spatially
separated entangled states in quantum mechanics. This may lay the
groundwork for a consistent quantum gravity theory.

\vskip 0.3 true in

{\bf Acknowledgments}
\vskip 0.2 true in
This work was supported by the Natural Sciences and Engineering Research Council of
Canada. I thank Michael Clayton and Pierre Savaria for stimulating and helpful
conversations.\vskip 0.5 true in

 \end{document}